\documentclass{iopjournal}
\usepackage{amsmath}
\usepackage{cite}

\begin{document}

\articletype{Paper} 

\title{Phase coding semi-quantum key distribution system based on the Single-state protocol}

\author{Si-Ying Huang$^1$, Qin-Cheng Hou$^1$, Tian-Ming Zhao$^{2,*}$, Jin-Dong Wang$^{1,*}$, Nai-Da Mo$^1$, Zheng-Jun Wei$^1$, Ya-Fei Yu$^2$ and Zhi-Ming Zhang$^3$}

\affil{$^1$Guangdong Provincial Key Laboratory of Quantum Engineering and Quantum Materials, School of Optoelectronic Science and Engineering, South China Normal University, Guangzhou 510006, China}

\affil{$^2$Guangdong Provincial Key Laboratory of Nanophotonic Functional Materials and Devices, School of Optoelectronic Science and Engineering, South China Normal University, Guangzhou 510006, China}

\affil{$^3$Center of Physics Experiments, Guangdong Technology College, Zhaoqing, 526100, China}

\email{zhaotm@scnu.edu.cn, wangjindong@m.scnu.edu.cn}


\begin{abstract}
Semi-quantum key distribution (SQKD) allows sharing random keys between a quantum user and a classical user, which significantly saves user resources, especially when using the Single-state protocol. However, the operation of the classical user, which involves measurement and resending using the Single-state protocol, presents technical difficulties in experiment and there is a security vulnerability of “tagged” attack in theory. To solve these problems, in our work, based on the Single-state protocol, we propose the “selective modulation" method and successfully implement a phase-encoded semi-quantum key distribution system. The system operates at a frequency of 100MHz and an average photon number of 0.1. The interference contrast achieved 97.45\%, the average quantum bit error rate was 1.20\%, and the raw key rate reached 88Kbps. Our experimental results demonstrate the feasibility and stability of the proposed phase-encoded SQKD system. Furthermore, we conducted an analysis of the “selective modulation" scheme in terms of quantum state evolution to assess the security of our system and ultimately proved that it can resist “tagged” attack. The classical user of our system requires only two optical devices and operates without relying on full quantum capabilities, thereby enhancing its application potential in quantum networks. This work validates the feasibility of SQKD experiments and provides ideas for future research on SQKD experiments and security studies.
\end{abstract}
 
\section{Introduction}
Quantum key distribution (QKD) allows the sharing of a string of random keys between legitimate users, with security guarantees provided by the principles of quantum mechanics: eavesdropping on the key distribution process will be detected. Currently, QKD has reached the stage of practical application, and its application over fiber and free space has been validated by numerous experiments\cite{RN95, RN473, RN474, RN475, RN476, RN477, RN478, RN94}. In 2007, Boyer et al. proposed the concept of semi-quantum key distribution (SQKD)\cite{RN149}, and this is called the BKM07 protocol. SQKD allows one party to perform only classical operations, thereby reducing the quantum resource requirement in the key distribution process. If SQKD can reduce the requirements for quantum resources, then it will have unique application prospects and value. Since 2007, various semi-quantum key distribution protocols have been proposed\cite{RN158, RN151, RN480, RN481, RN482, RN228, RN483, RN150}. In 2009, Boyer et al. further based on randomized SQKD protocol and measuring retransmitting the SQKD, proved their robustness\cite{RN158}, and Zou et al. proposed the Single-state protocol \cite{RN150}. The Single-state protocol prepares only one quantum state, which holds a higher application value. Subsequently, the semi-quantum model expanded into areas such as quantum secret sharing (QSS)\cite{RN484, RN485}, quantum secure direct communication (QSDC)\cite{RN486}, quantum key agreement (QKA)\cite{RN487}, and quantum authentication\cite{RN488}, evolving into semi-quantum cryptography. Regarding the security of SQKD, Boyer et al. introduced the concept of robustness\cite{RN149}. In 2014, Krawec et al. pointed out that collective attacks in single-state SQKD can be regarded as restricted attacks\cite{RN489}. In 2018, the security of Single-state SQKD protocols was proven\cite{RN490}. In 2022, Mi et al. conducted a security analysis on the Photon-Number Splitting (PNS) attack problem caused by multi-photon pulses in semi-quantum protocols\cite{RN225}. In 2023, Dong et al. proposed a decoy state construction scheme for the four-state SQKD protocol\cite{RN227}.

The theoretical research of SQKD has been continuously studied since BKM07 was proposed. However, the experimental research of SQKD has faced a great challenge. SQKD requires the classical party to perform two types of classical operations: CTRL, which directly sends back photons, and SIFT, which measures and resends photons. Implementing the SIFT operation is difficult and vulnerable to the “tagged” attack, which can distinguish classical operations. This challenge has led to slow progress in SQKD experiments. In 2012, Boyer et al. proposed the Mirror protocol\cite{RN151}. In 2021, Han et al. used time-phase encoding and the “selective modulation” method to implement the classical party's operations in the Mirror protocol, completing the first principal demonstration of SQKD\cite{RN120}. The demonstration of the Mirror protocol prompted us to consider whether the “selective modulation” method could be applied to other protocols, such as the Single-state protocol, and to provide a security analysis of the method. In 2024, Mo et al. completed the principle verification of Single-state protocol on the free space channel by polarization dimension, and the security of “selective modulation" is proved\cite{RN224}. However, the Single-state protocol on the fiber channel has not yet been verified. Therefore, we developed a phase-encoded optical system based on the Single-state protocol and verified its feasibility using the “selective modulation" method. Our system avoids the problem of measuring and retransmitting photons, maintains good stability, and improves the utilization efficiency of photons. To evaluate the security of the system, we conducted a security analysis of the "selective modulation" method in the system scheme and proved that it is resistant to “tagged” attack. Our system represents a new application of the selective modulation method in SQKD. Moreover, the security analysis has for the first time demonstrated the security of the "selective modulation" method.

The remainder of this paper is organized as follows: In Sec.\ref{sec:2}, the Single-state protocol is briefly introduced. In Sec.\ref{sec:3}, we propose the phase encoding semi-quantum key distribution based on the Single-state protocol, and then the experiment and results are described in detail in Sec.\ref{sec:4}. In Sec.\ref{sec:5}, we conducted a security analysis of the selective modulation used in this paper. We conclude in Sec.\ref{sec:6}.

\section{The Single-state protocol}\label{sec:2}
Since this paper focuses on the Single-state protocol, we briefly introduce it.  In the Single-state protocol, Bob only needs to prepare one state, which greatly reduces the preparation difficulty of the quantum user. The protocol process is as follows:
Bob prepares a quantum bit in the $\left|  +  \right\rangle $ state and sends it to Alice. Alice randomly chooses one of two classical operations upon receiving the quantum bit: (1) CTRL: returns the received quantum bit directly; (2) SIFT: measures the received quantum bit in Z basis, prepares the same quantum bit as her measured result, and sends it back to Bob. Bob randomly chooses to measure the returned quantum bit in either the Z basis or the X basis. Bob announces his measurement basis choice, and Alice announces her operation. As a result: If Alice chooses SIFT and Bob chooses Z-basis measurement, the resulting bits are named SIFT-Z bit. These bits contain information exchanged between Alice and Bob and can be used as the raw key. If Alice chooses CTRL and Bob chooses X-basis measurement, the resulting bits are named CTRL-X bits. Bob checks the error rate of the CTRL-X bits, and if the error rate exceeds the set threshold, the protocol is terminated. Furthermore, Alice and Bob choose a portion of the SIFT-Z bits to check the error rate. If the error rate of this portion of bits is higher than the set threshold, the protocol is terminated. Finally, Alice and Bob use the remaining SIFT-Z bits as information bits (INFO bits).

Although the Single-state protocol can reduce the difficulty of state preparation, it still requires Alice to measure and resend photons, making it susceptible to the “tagged” attack\cite{RN491} and difficult to achieve in the experiment. The protocol process indicates that measurements at Alice are used only for randomly generating information bits, rather than for eavesdropping detection. This leads us to consider whether the “selective modulation” method can be employed in the Single-state protocol to mitigate the security concerns arising from measuring and resending photons. “Selective modulation"\cite{RN120, RN224} means that instead of using measurement and resending for SIFT operation in the classical user, the state is directly modulated, and CTRL is still directly returned. The work on this paper is based on the above considerations.

\section{The phase encoding semi-quantum key distribution based on the Single-state protocol}\label{sec:3}
\begin{figure}[ht]
	\centering
	\includegraphics[width=0.5\textwidth]{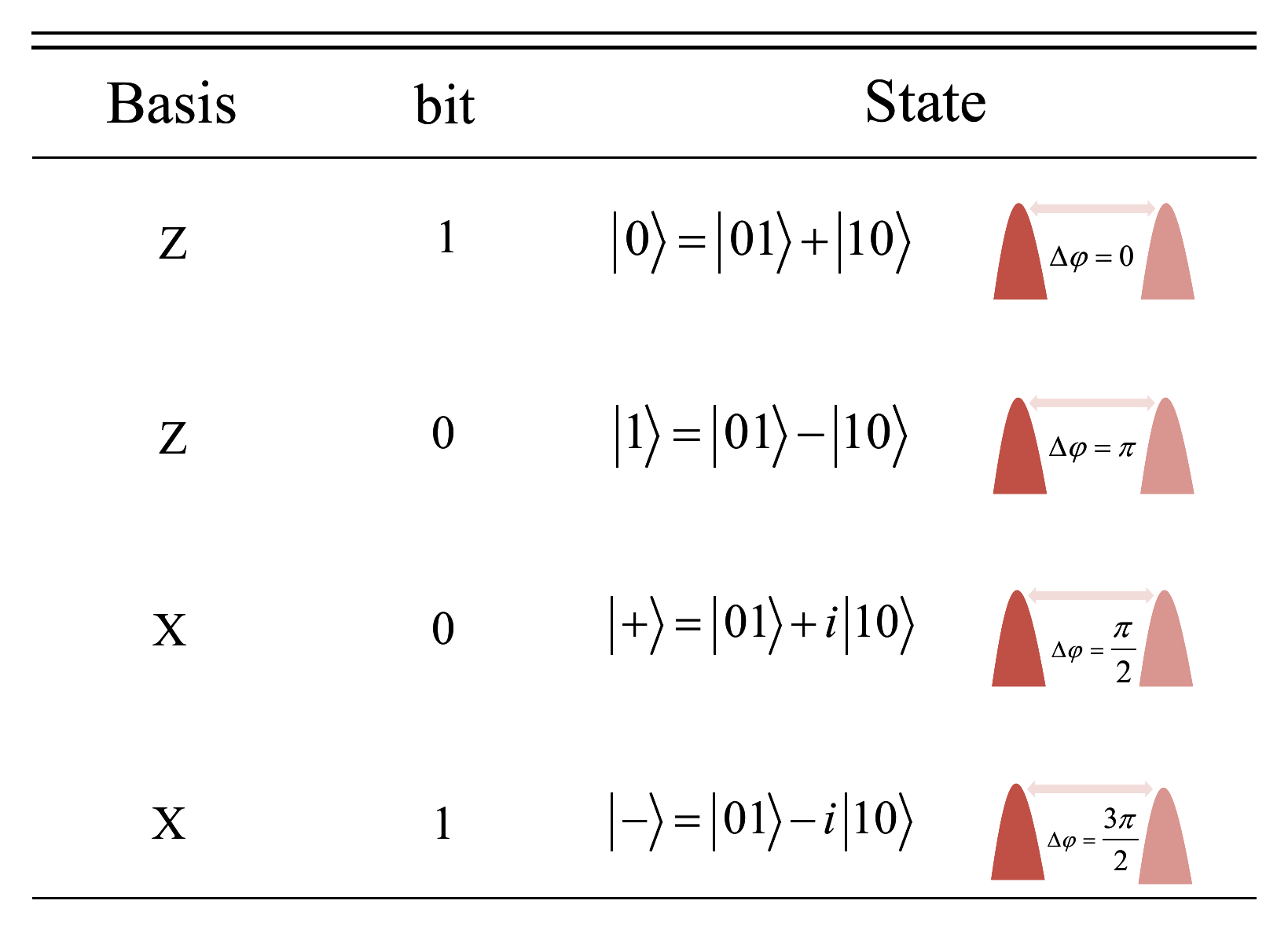}
	\caption{Phase encoding of the states.}
	\label{fig:figure1}
\end{figure}

Based on the Single-state protocol, our SQKD system employs phase encoding, in which both encoding and decoding are achieved by modulating the phase difference between two pulses.  Here, we define Z basis as the phase difference between two pulses being $0$ or $\pi $. Conversely, we define X basis as the phase difference between two pulses being $\frac{\pi }{2}$ or $\frac{{3\pi }}{2}$. Quantum states under each basis are illustrated in Fig. \ref{fig:figure1}, where $\left| {01} \right\rangle $ and $\left| {10} \right\rangle $ represent photons in distinct time bins under the Fock state.

\begin{figure}[ht]
	\centering
	\includegraphics[width=0.7\linewidth]{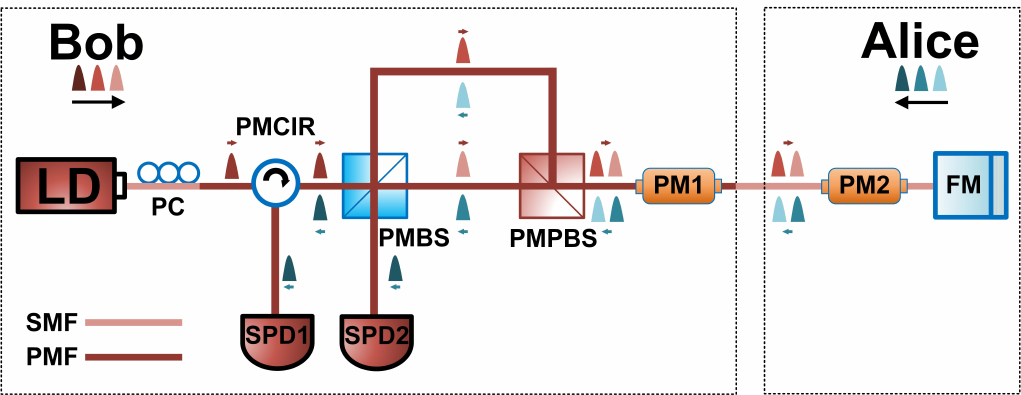}
	\caption{The scheme of phase encoding semi-quantum key distribution. LD: laser diode, PC: polarization controller, PMCIR: polarization maintaining circulator, PMBS: polarization maintaining beam splitter, PMPBS: polarization maintaining polarization beam splitter, PMs(PM1 and PM2): phase modulator, FM: faraday mirror, SPDs (SPD1 and SPD2): single-photon detectors, SMF: single mode fiber, PMF: polarization maintaining fiber. The reddish-brown envelope indicates the forward-transmitted light pulse, that is, from Bob to Alice. The blue-green envelope indicates the backward-transmitted light pulse, that is, from Alice back to Bob.}
	\label{fig:figure2}
\end{figure}

\begin{figure}[ht]
	\centering
	\includegraphics[width=12cm]{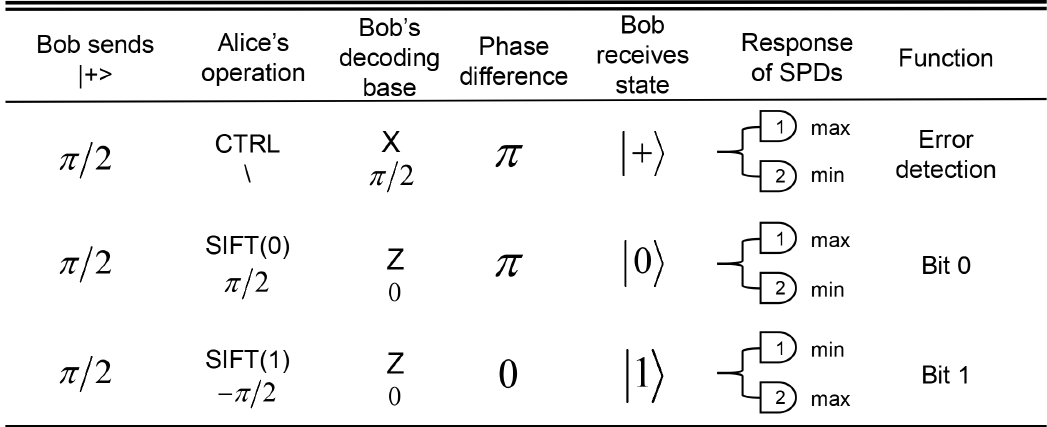}
	\caption{The relationship between Alice and Bob's operations and the response of the single photon detector. $\begin{array}{*{20}{c}}
			{{\rm{CTRL }}}\\
			\backslash 
		\end{array}$ indicates that Alice selects CTRL operation, while PM does not apply voltage. $\begin{array}{*{20}{c}}
			{{\rm{SIFT}}\left( {\rm{0}} \right)}\\
			{{\pi  \mathord{\left/
						{\vphantom {\pi  2}} \right.
						\kern-\nulldelimiterspace} 2}}
		\end{array}$ indicates that Alice selects SIFT(0) operation, by changing the phase ${\pi  \mathord{\left/
			{\vphantom {\pi  2}} \right.
			\kern-\nulldelimiterspace} 2}$ on PM. $\begin{array}{*{20}{c}}
			{\rm{X}}\\
			{{\pi  \mathord{\left/
						{\vphantom {\pi  2}} \right.
						\kern-\nulldelimiterspace} 2}}
		\end{array}$ indicates that Bob selects X basis,  by changing the phase ${\pi  \mathord{\left/
				{\vphantom {\pi  2}} \right.
				\kern-\nulldelimiterspace} 2}$ on PM.
	}
	\label{fig:figure3}
\end{figure}

The scheme of phase encoding semi-quantum key distribution is illustrated in Fig. \ref{fig:figure2}. Bob prepares and sends the quantum bit $\left|  +  \right\rangle $ by using an asymmetric Mach-Zehnder interferometer (AMZI) and PM1. Upon receiving the quantum bit, Alice randomly selects one of two classical operations: CTRL: directly returns the received quantum bit; SIFT: selectively modulates the received quantum bit by PM2. For SIFT(0), Alice applies $\frac{\pi }{2}$ phase modulation, transforming the quantum state to $\left| 0 \right\rangle $. For SIFT(1), Alice applies $ - \frac{\pi }{2}$ phase modulation, transforming the quantum state to $\left| 1 \right\rangle $. After selective modulation, Alice sends the quantum bit back to Bob. Bob randomly chooses to measure the quantum bit using either the X basis or the Z basis. When measuring in the X basis, Bob's phase modulator applies either $\frac{\pi }{2}$ or $ - \frac{\pi }{2}$ phase modulation. When measuring in the Z basis, Bob set the voltage of the phase modulator to 0. When the photon reaches the polarization maintaining beam splitter (PMBS), interference occurs. Then the photon will enter different single-photon detectors based on the phase difference. The relationship between the response of the single-photon detectors and the operation of both Alice and Bob is depicted in Fig. \ref{fig:figure3}. Only the cases where Alice's chosen operation and Bob's chosen basis are compatible are presented here.

\section{Experiment and Results}\label{sec:4}

\subsection{Experimental setup}\label{4.1}
\begin{figure}[ht]
	\centering
	\includegraphics[width=0.7\linewidth]{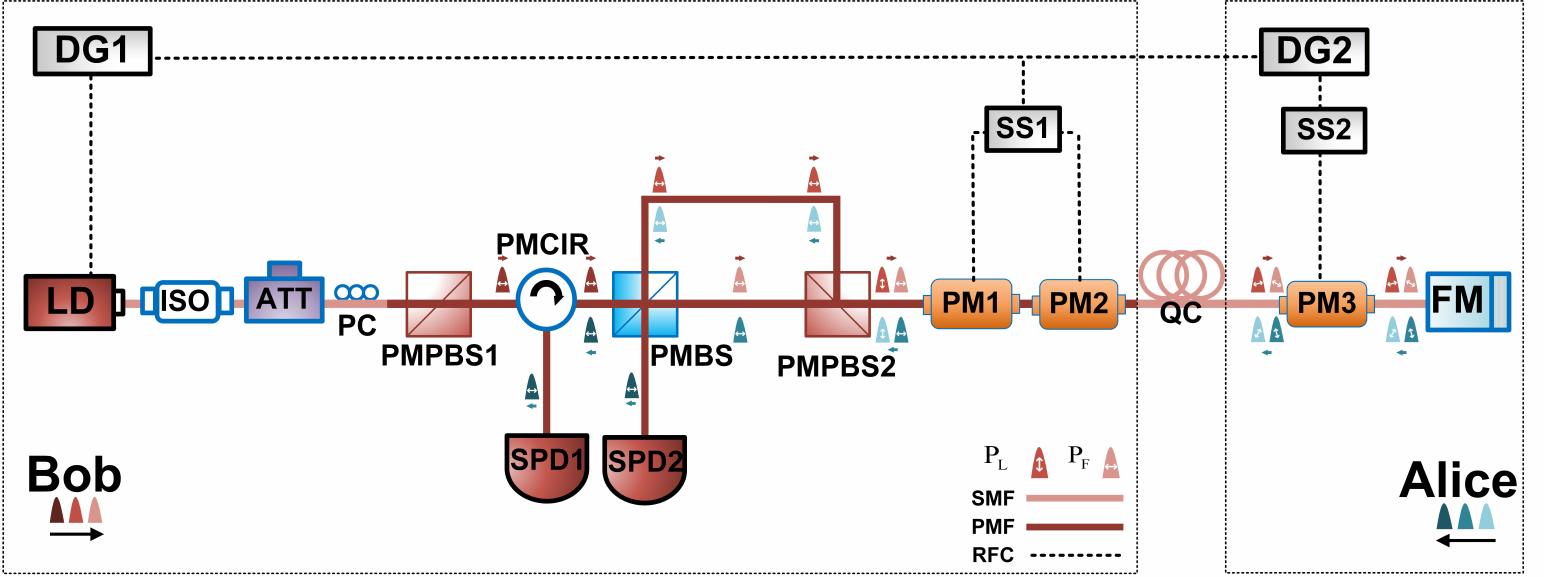}
	\caption{Experimental setup of phase encoding semi-quantum key distribution based on Single-state protocol. ATT: attenuator, QC: quantum channel, DG: digital generator, SS: signal source, RFC: radio frequency cable.}
	\label{fig:figure4}
\end{figure}

According to the previously described scheme, the experimental setup in this article is illustrated in Fig. \ref{fig:figure4}. On Bob's module, the setup comprises four modules: a laser source, an AMZI, phase modulation, and detectors. In the source section, light pulses are generated by a picosecond laser diode (LD). The pulse width of the light pulses is 50ps, and the system frequency is 100MHz. The PC and the PMPBS1 are used to adjust the polarization state of the light pulses and align them with the slow axis of the polarization-maintaining fiber. In the AMZI section, the PMCIR is used to transmit the forward light pulses to the 50:50 PMBS and separate the backward light pulses to the single-photon detectors (SPDs). The PMBS, along with PMPBS2, forms the AMZI. The phase modulation section consists of two polarization-independent phase modulators (PM1, PM2), which modulate the forward and backward light pulses. A signal source (SS1) drives the phase modulators. The detection frequency of SPDs is 100 MHz. The clock synchronization of the LD, SPDs, and SS1 is achieved by a digital generator (DG).

On Alice's module, the setup consists of a polarization-independent phase modulator (PM3) and a Faraday mirror. The phase modulator is driven by SS2.

Based on the previous description, to implement classical operations in SQKD, we employ a “selective modulation” method in our scheme instead of the “original measure-and-resend” method. In the AMZI, we introduce a delay of 2.9ns between the two pulses for the arm length difference. The two light pulses are separated by AMZI and propagate along the fast axis and slow axis of the polarization-maintaining fiber respectively.  For simplicity, we will refer to them as ${P_F}$ and ${P_L}$. When the pulses arrive at Alice, Alice selectively modulates ${P_L}$ using her phase modulator (PM3). When the pulses are reflected back to Bob, Bob modulates the pulses using his phase modulator (PM2). Due to the Faraday rotation conjugate effect, the polarization of pulse will rotate by $90^{\circ}$. Consequently, both ${P_F}$ and ${P_L}$ will return to the PMBS simultaneously and undergo interference. Finally, depending on the phase difference, the pulses will be detected by either SPD1 or SPD2.

The “selective modulation” operation is described in detail below: Bob sends light pulses. A pulse signal with a repetition frequency of 100MHz generated by SS1, drives the phase modulator PM1. Then the PM1 consistently applies $\frac{\pi }{2}$ phase modulation on ${P_L}$. If Alice chooses CTRL, she does not drive the phase modulator, and the light pulse will be reflected to Bob directly. If Alice chooses SIFT(0), she applies $\frac{\pi }{2}$ phase modulation on ${P_L}$. In this situation, PM3 is driven by a pulse signal with a repetition frequency of 100MHz, generated from SS2. If Alice chooses SIFT(1), she applies $ - \frac{\pi }{2}$ phase modulation on ${P_L}$, achieved by changing the signal polarity. In order not to disrupt the Faraday effect, Alice needs to modulate the forward and backward pulses in the same way twice. This can be achieved by controlling the length of the fiber between PM3 and FM.

When the pulses return to Bob, he randomly chooses to measure them in either the Z basis or the X basis. If Bob chooses the Z basis measurement, he drives PM2 with a voltage of 0. If Bob chooses the X measurement, he applies $\frac{\pi }{2}$ phase modulation to ${P_F}$. In this case, PM2 is driven by a pulse signal with a repetition frequency of 100MHz, generated by SS2.  Here, in order to avoid the inconsistency in the detection efficiency and to balance the response counts of the two detectors, we further perform $ - \frac{\pi }{2}$ phase modulation on ${P_f}$. In our system, Bob employs two phase modulators to achieve encoding and decoding functions, aiming to reduce modulation complexity. In future practical applications, through optimizing system design and improving the accuracy of electrical signals, it will be possible to achieve the aforementioned functions with only one phase modulator.

If Alice chooses the SIFT(0) operation and Bob chooses the Z basis measurement, the light pulse will reverse into PMCIR and then into SPD1 because of the interference of PMBS. If Alice chooses the SIFT(1) operation and Bob chooses the Z basis measurement, the light pulse will enter SPD2. If Alice chooses the CTRL operation and Bob chooses the X measurement, the pulse will enter SPD2. 

\subsection{Results and discussion}\label{4.2}
\begin{figure}[ht]
	\centering
	\includegraphics[width=0.7\textwidth]{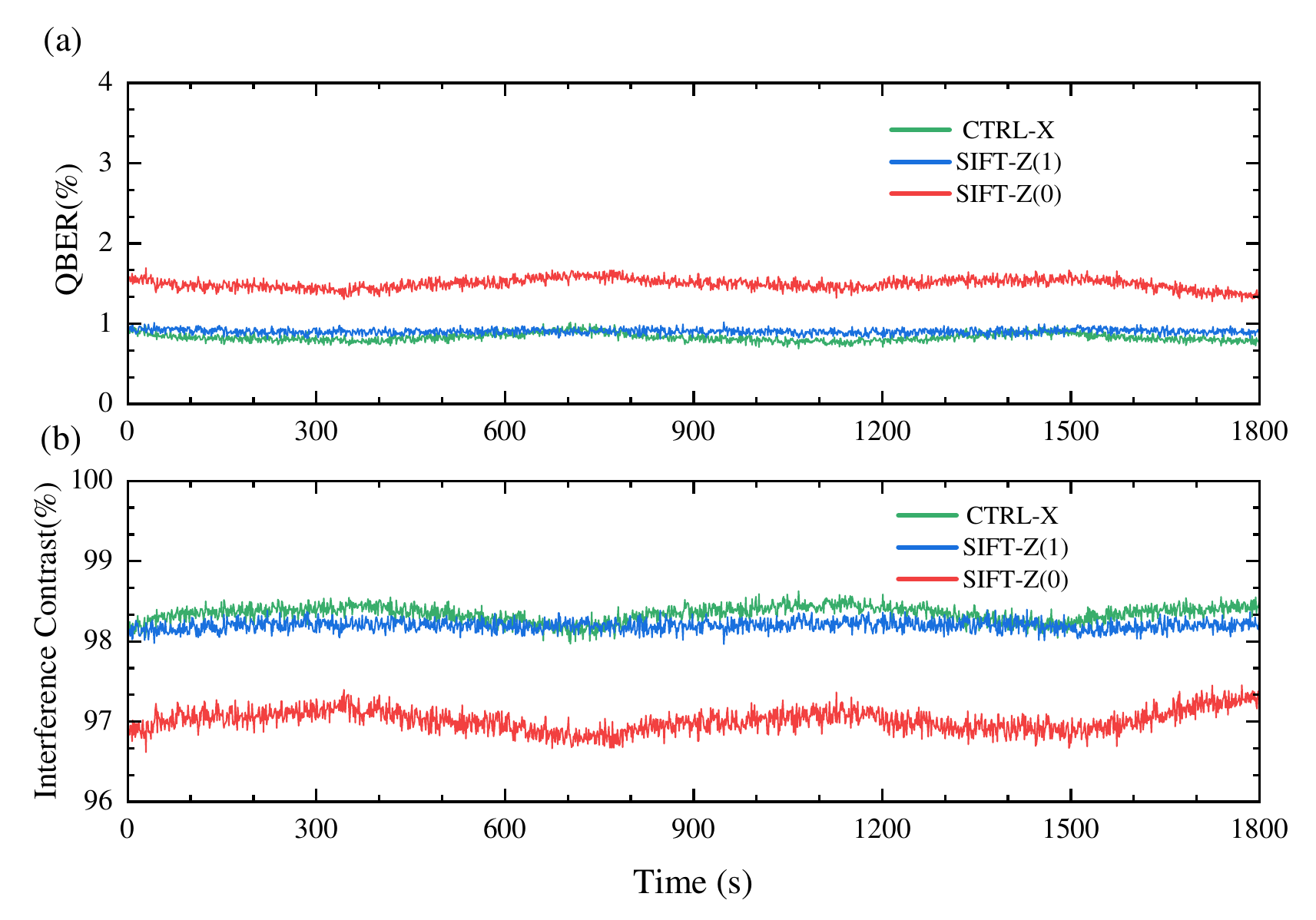}
	\caption{(a) The change of error rate within 30 minutes. (b) The change of interference contrast within 30 minutes.}
	\label{fig:figure5}
\end{figure}
According to the system design and experimental settings described earlier, to evaluate the system performance, we collected the response results of the SPDs over a 30-minute period. The overall error rate and interference contrast of each base within 30 minutes are as follows: CTRL-X: average error rate is 0.82\%, interference contrast is 98.35\%, and average response rate is 0.92\%; SIFT-Z: average error rate is 1.20\%, interference contrast is 97.45\%, and average response rate is 0.90\%. The change of error rate within 30 minutes is shown in Fig. \ref{fig:figure5}(a), and the change of interference contrast is shown in Fig. \ref{fig:figure5}(b). Overall, it is evident that the interference contrast and error rate of our system are maintained at a stable level. With an average number of photons per pulse of 0.1, the raw key rate reaches 88Kbps, representing a significant improvement over the other experimental schemes\cite{RN120, RN224} in the field of semi-quantum key distribution. The experimental results validate the feasibility and stability of the phase encoding semi-quantum key distribution based on the Single-state protocol.

Notably, when the phase difference between light pulses is $\pi $, the response rate, error rate, and interference contrast have a slight decline compared to when the phase difference is 0. This discrepancy arises because at a phase difference of $\pi$, photons will be output from port 3 of the PMCIR, while there is a 0.39 dB attenuation from port 2 to port 3. Therefore, although detectors with similar detection efficiencies were selected for our system, the attenuation introduced by the PMCIR results in certain discrepancies in the output results at both ends, resulting in an increase in overall error rate.

\section{Security proof}\label{sec:5}

\begin{figure}[t]
	\centering
	\includegraphics[width=0.5\textwidth]{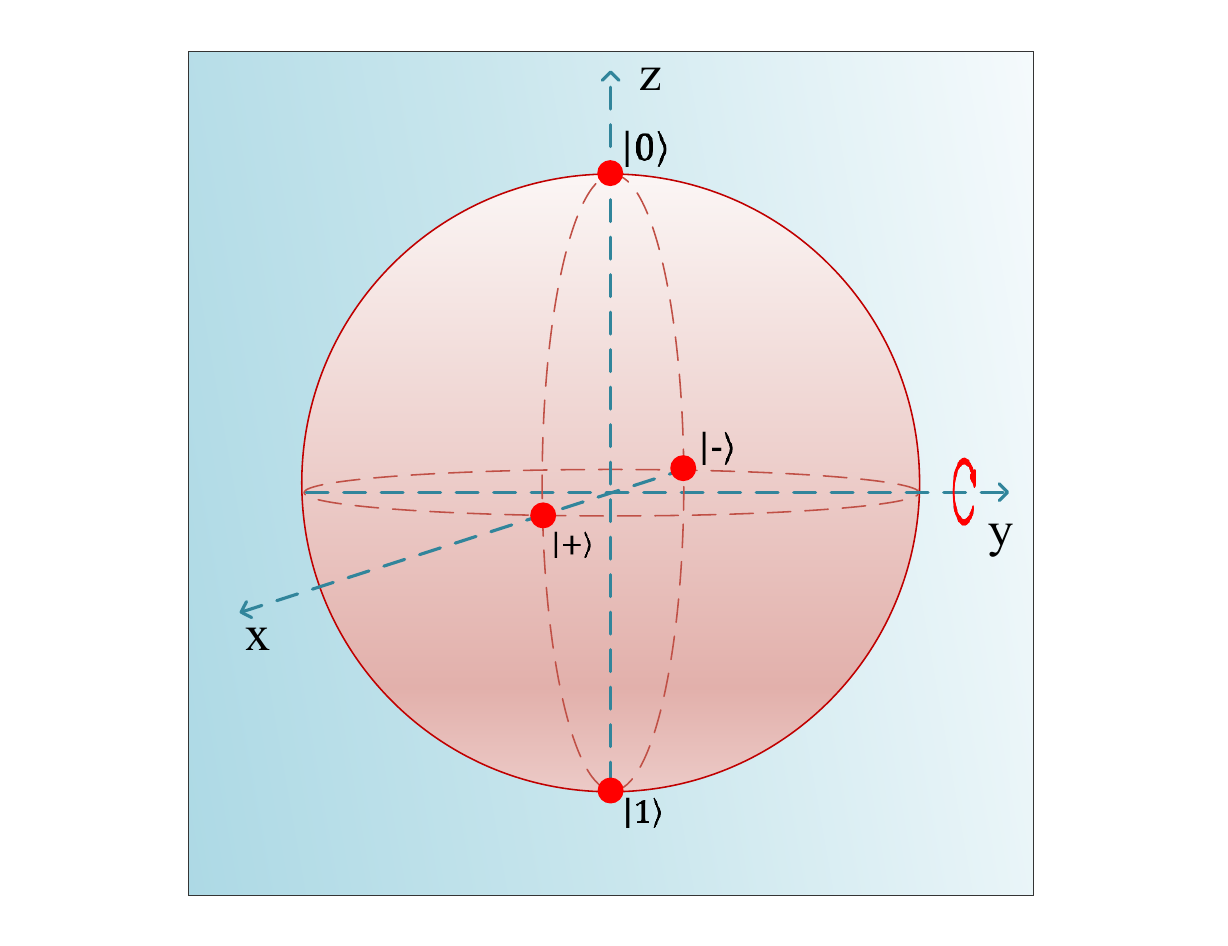}
	\caption{Quantum states on the Bloch sphere.}
	\label{fig:figure6}
\end{figure}

A security analysis is essential as this paper employs “selective modulation” instead of the original measurement-resend operation. Here, we demonstrate that if Eve attacks quantum states to obtain key information during the distribution process, errors will inevitably arise and be detected by Alice and Bob. Initially, Bob prepares the qubit state ${\left| + \right\rangle _B}$ and sends them one by one to Alice, sending the next quantum bit only after receiving the photon. As depicted in Fig. \ref{fig:figure1}, ${\left| + \right\rangle _B}$ can be represented as:
\begin{align}
	{\left|  +  \right\rangle _B} = \frac{{1 + i}}{2}{\left| 0 \right\rangle _B} + \frac{{1 - i}}{2}{\left| 1 \right\rangle _B}.
	\label{<1>}%
\end{align}
It means the ${\left| + \right\rangle _B}$, ${\left| 0 \right\rangle _B}$ and ${\left| 1 \right\rangle _B}$ can be defined as poles on the equator of Bloch sphere (as shown in Fig. \ref{fig:figure6}). To describe more clearly and without loss of generality, we rewrite Eq. \eqref{<1>} as follows:

\begin{align}
	\setlength{\abovedisplayskip}{3pt}
	\setlength{\belowdisplayskip}{3pt}
	{\left|  +  \right\rangle _B} = \frac{1}{{\sqrt 2 }}\left( {{{\left| 0 \right\rangle }_B} + {{\left| 1 \right\rangle }_B}} \right),
	\label{<2>}%
\end{align}
which does not change the relationships between each state and the subsequent analysis results.

Now, we are modeling the “selective modulation” method. When Alice performs selective modulation on the received quantum state, choosing the CTRL operation is equivalent to applying the identity operator ${\hat I_A}$ to the quantum state. When Alice chooses the SIFT operation, she randomly rotates $\left|  +  \right\rangle $ to either $\left| 0 \right\rangle $ or $\left| 1 \right\rangle $, corresponding to rotating the quantum state around the y-axis of the Bloch sphere counterclockwise by $90\,^{\circ}$ or clockwise by $90\,^{\circ}$ (Fig. \ref{fig:figure6}). Therefore, the selective modulation process in this paper can be fully represented by a unitary operator corresponding to rotation around the y-axis, as shown in Eq. \eqref{<3>}, where $\delta $ represents the angle of counterclockwise rotation around the y-axis for the quantum state:
\begin{align}
	{\hat R_y}\left( \delta  \right) = \left[ {\begin{array}{*{20}{c}}
			{\cos \frac{\delta }{2}}&{ - \sin \frac{\delta }{2}}\\[6pt]
			{\sin \frac{\delta }{2}}&{\cos \frac{\delta }{2}}
	\end{array}} \right].
	\label{<3>}%
\end{align}

Because the classic user Alice can only perform quantum state operations on the Z basis, and only uses CTRL to return the state to the quantum user Bob in the X basis, therefore, here we define the range $\delta  \in \left\{ { - \frac{\pi }{2},\frac{\pi }{2}} \right\}$. Thus, the selective modulation process for any quantum state in this system can be represented as: 
\begin{align}
	\begin{array}{c}
		{{\hat S}_0} = {{\hat R}_y}\left( { - \frac{\pi }{2}} \right) ,
		\label{<4>}%
	\end{array}
\end{align}
\begin{align}
	\begin{array}{c}
		{{\hat S}_1} = {{\hat R}_y}\left( {\frac{\pi }{2}} \right),
		\label{<4'>}%
	\end{array}
\end{align}
Where ${\hat S_0}$ represents the SIFT(0) operation, and ${\hat S_1}$ represents the SIFT(1) operation. 
Note that the unitary operator in Eq.\eqref{<3>} represents an equivalent input--output map of the SIFT process, rather than a physically implemented quantum gate by Alice. Hadamard operation realizes the mapping of bases $H\left| 0 \right\rangle  = \left|  +  \right\rangle $, $H\left| 1 \right\rangle  = \left|  -  \right\rangle $. But the selective modulation is ${\hat R_y}\left( \delta  \right)$, $\delta  \in \left\{ { - \frac{\pi }{2}, \frac{\pi }{2}} \right\}$. Therefore, ${\hat R_y}\left( { \pm \frac{\pi }{2}} \right)\left|  +  \right\rangle  = \left| 0 \right\rangle \ \text{or}\ \left| 1 \right\rangle$. Alice has never mapped the Z-basis state back to the X-basis state and lacks the ability to generate $\left|  -  \right\rangle $ thus does not possess the complete BB84 preparation capability. So, we define Alice as a classical user.

To attack quantum states, Eve use two unitary operators ${\hat U_F}$ and ${\hat U_R}$ to perform attack on ${\left| \phi  \right\rangle _B}$. ${\hat U_F}$ is acts on the ${\left| \phi  \right\rangle _B}$ in forward channel and Eve's ancilla state ${\left| \chi  \right\rangle _E}$ (${\left| \chi  \right\rangle _E} \in {{\cal H}_E}$,${{\cal H}_E}$ is the two-dimensional Hilbert space corresponding to Eve's probe state). ${\hat U_R}$ is acts on the ${\left| \phi  \right\rangle _B}$ in backwards channel and Eve's ancilla state ${\left| \chi  \right\rangle _E}$. In forward channel, the attack ${\hat U_F}$ can be defined as follow:
\begin{align}
	\begin{array}{l}
		{{\hat U}_F}\left( {{{\left| 0 \right\rangle }_B}{{\left| \chi  \right\rangle }_E}} \right) = {\left| 0 \right\rangle _B}\left| {{\chi _{00}}} \right\rangle  + {\left| 1 \right\rangle _B}\left| {{\chi _{01}}} \right\rangle ,
	\end{array}
	\label{<5>}%
\end{align}
\begin{align}
	\begin{array}{l}
		{{\hat U}_F}\left( {{{\left| 1 \right\rangle }_B}{{\left| \chi  \right\rangle }_E}} \right) = {\left| 0 \right\rangle _B}\left| {{\chi _{10}}} \right\rangle  + {\left| 1 \right\rangle _B}\left| {{\chi _{11}}} \right\rangle.
	\end{array}
	\label{<5'>}%
\end{align}
Therefore, after executing the attack ${\hat U_F}$, the states held by Bob and Eve can be represented as:
\begin{align}
	\begin{array}{c}
		{{\hat U}_F}\left( {{{\left|  +  \right\rangle }_B}{{\left| \chi  \right\rangle }_E}} \right)
		= {{\hat U}_F}\left[ {\frac{1}{{\sqrt 2 }}\left( {{{\left| 0 \right\rangle }_B} + {{\left| 1 \right\rangle }_B}} \right){{\left| \chi  \right\rangle }_E}} \right]\\[2mm]
		= \frac{1}{{\sqrt 2 }}\left( {{{\left| 0 \right\rangle }_B}\left| {{\chi _{00}}} \right\rangle  + {{\left| 1 \right\rangle }_B}\left| {{\chi _{01}}} \right\rangle } \right.
		\left. { + {{\left| 0 \right\rangle }_B}\left| {{\chi _{10}}} \right\rangle  + {{\left| 1 \right\rangle }_B}\left| {{\chi _{11}}} \right\rangle } \right).
	\end{array}
	\label{<6>}%
\end{align}

When Alice chooses the CTRL operation, the states held by Bob and Eve can still be represented as Eq.\eqref{<6>}. When Alice chooses the SIFT(0) operation, the states held by Alice, Bob, and Eve can be represented as:
\begin{align}
	\begin{array}{c}
		{\left| \psi  \right\rangle _{{A_0}B}}{\left| \chi  \right\rangle _E} = {\left| 0 \right\rangle _A} \otimes {{\hat S}_0}{{\hat U}_F}\left( {{{\left|  +  \right\rangle }_B}{{\left| \chi  \right\rangle }_E}} \right)\\[2mm]
		= \frac{1}{2}\left( {{{\left| {00} \right\rangle }_{AB}}\left| {{\chi _{00}}} \right\rangle  - {{\left| {01} \right\rangle }_{AB}}\left| {{\chi _{00}}} \right\rangle  + {{\left| {00} \right\rangle }_{AB}}\left| {{\chi _{01}}} \right\rangle } \right.
		+ {\left| {01} \right\rangle _{AB}}\left| {{\chi _{01}}} \right\rangle  + {\left| {00} \right\rangle _{AB}}\left| {{\chi _{10}}} \right\rangle  - {\left| {01} \right\rangle _{AB}}\left| {{\chi _{10}}} \right\rangle \\[2mm]
		\left. { + {{\left| {00} \right\rangle }_{AB}}\left| {{\chi _{11}}} \right\rangle  + {{\left| {01} \right\rangle }_{AB}}\left| {{\chi _{11}}} \right\rangle } \right).
	\end{array}
	\label{<7>}%
\end{align}
When Alice chooses the SIFT(1) operation, the states held by Alice, Bob, and Eve can be represented as:
\begin{align}
	\begin{array}{c}
		{\left| \psi  \right\rangle _{{A_1}B}}{\left| \chi  \right\rangle _E} = {\left| 1 \right\rangle _A} \otimes {{\hat S}_1}{{\hat U}_F}\left( {{{\left|  +  \right\rangle }_B}{{\left| \chi  \right\rangle }_E}} \right)\\[2mm]
		= \frac{1}{2}\left( {{{\left| {10} \right\rangle }_{AB}}\left| {{\chi _{00}}} \right\rangle  + {{\left| {11} \right\rangle }_{AB}}\left| {{\chi _{00}}} \right\rangle  - {{\left| {10} \right\rangle }_{AB}}\left| {{\chi _{01}}} \right\rangle } \right.
		+ {\left| {11} \right\rangle _{AB}}\left| {{\chi _{01}}} \right\rangle  + {\left| {10} \right\rangle _{AB}}\left| {{\chi _{10}}} \right\rangle  + {\left| {11} \right\rangle _{AB}}\left| {{\chi _{10}}} \right\rangle \\[2mm]
		\left. { - {{\left| {10} \right\rangle }_{AB}}\left| {{\chi _{11}}} \right\rangle  + {{\left| {11} \right\rangle }_{AB}}\left| {{\chi _{11}}} \right\rangle } \right).
	\end{array}
	\label{<8>}%
\end{align}

After Alice performs selective modulation, Eve implements an attack ${\hat U_R}$ on the quantum state of the backward channel. The final quantum state of Alice 's chosen CTRL is: 
\begin{align}
	\begin{array}{c}
		{\left| \psi  \right\rangle _B}{\left| {{\chi _{final}}} \right\rangle _E} = {{\hat U}_R}{{\hat U}_F}\left( {{{\left|  +  \right\rangle }_B}{{\left| \chi  \right\rangle }_E}} \right)\\[2mm]
		= \frac{1}{{\sqrt 2 }}\left( {{{\left| 0 \right\rangle }_B}\left| {{\chi _{0000}}} \right\rangle  + {{\left| 1 \right\rangle }_B}\left| {{\chi _{0001}}} \right\rangle } \right.
		+ {\left| 0 \right\rangle _B}\left| {{\chi _{0110}}} \right\rangle  + {\left| 1 \right\rangle _B}\left| {{\chi _{0111}}} \right\rangle  + {\left| 0 \right\rangle _B}\left| {{\chi _{1000}}} \right\rangle \\[2mm]
		\left. { + {{\left| 1 \right\rangle }_B}\left| {{\chi _{1001}}} \right\rangle  + {{\left| 0 \right\rangle }_B}\left| {{\chi _{1110}}} \right\rangle  + {{\left| 1 \right\rangle }_B}\left| {{\chi _{1111}}} \right\rangle } \right).
	\end{array}
	\label{<9>}%
\end{align}

If Eve introduces no errors on the CTRL-X bit, which means the measurement result state in Bob could not be ${\left| -  \right\rangle }$ , there should be:
\begin{align}
	\begin{array}{c}
		\left| {{\chi _{0000}}} \right\rangle  + \left| {{\chi _{0110}}} \right\rangle  + \left| {{\chi _{1000}}} \right\rangle  + \left| {{\chi _{1110}}} \right\rangle 
		= \left| {{\chi _{0001}}} \right\rangle  + \left| {{\chi _{0111}}} \right\rangle  + \left| {{\chi _{1001}}} \right\rangle  + \left| {{\chi _{1111}}} \right\rangle.
	\end{array}
	\label{<10>}%
\end{align}

The final quantum state of Alice 's chosen SIFT(0) is:

\begin{align}
	\begin{array}{c}
		{{\hat U}_R}\left( {{{\left| \psi  \right\rangle }_{{A_0}B}}{{\left| \chi  \right\rangle }_E}} \right) = \frac{1}{2}\left( {{{\left| {00} \right\rangle }_{AB}}\left| {{\chi _{0000}}} \right\rangle } \right. + {\left| {01} \right\rangle _{AB}}\left| {{\chi _{0001}}} \right\rangle  - {\left| {00} \right\rangle _{AB}}\left| {{\chi _{0010}}} \right\rangle  - {\left| {01} \right\rangle _{AB}}\left| {{\chi _{0011}}} \right\rangle \\[2mm]
		+ {\left| {00} \right\rangle _{AB}}\left| {{\chi _{0100}}} \right\rangle  + {\left| {01} \right\rangle _{AB}}\left| {{\chi _{0101}}} \right\rangle 
		+ {\left| {00} \right\rangle _{AB}}\left| {{\chi _{0110}}} \right\rangle  + {\left| {01} \right\rangle _{AB}}\left| {{\chi _{0111}}} \right\rangle  
		+ {\left| {00} \right\rangle _{BA}}\left| {{\chi _{1000}}} \right\rangle  + {\left| {01} \right\rangle _{BA}}\left| {{\chi _{1001}}} \right\rangle \\[2mm]
		- {\left| {00} \right\rangle _{BA}}\left| {{\chi _{1010}}} \right\rangle  - {\left| {01} \right\rangle _{BA}}\left| {{\chi _{1011}}} \right\rangle 
		+ {\left| {00} \right\rangle _{BA}}\left| {{\chi _{1100}}} \right\rangle  + {\left| {01} \right\rangle _{BA}}\left| {{\chi _{1101}}} \right\rangle 
		\left. { + {{\left| {00} \right\rangle }_{BA}}\left| {{\chi _{1110}}} \right\rangle  + {{\left| {01} \right\rangle }_{BA}}\left| {{\chi _{1111}}} \right\rangle } \right).
		\label{<11>}%
	\end{array}
\end{align}

The final quantum state of Alice 's chosen SIFT(1) is:

\begin{align}
	\begin{array}{c}
		{{\hat U}_R}\left( {{{\left| \psi  \right\rangle }_{{A_1}B}}{{\left| \chi  \right\rangle }_E}} \right) = \frac{1}{2}\left( {{{\left| {10} \right\rangle }_{AB}}\left| {{\chi _{0000}}} \right\rangle  + {{\left| {11} \right\rangle }_{AB}}\left| {{\chi _{0001}}} \right\rangle } \right. 
		+ {{\left| {10} \right\rangle }_{AB}}\left| {{\chi _{0010}}} \right\rangle  + {{\left| {11} \right\rangle }_{AB}}\left| {{\chi _{0011}}} \right\rangle \\[2mm]
		- {{\left| {10} \right\rangle }_{AB}}\left| {{\chi _{0100}}} \right\rangle  - {{\left| {11} \right\rangle }_{AB}}\left| {{\chi _{0101}}} \right\rangle  
		+ {{\left| {10} \right\rangle }_{AB}}\left| {{\chi _{0110}}} \right\rangle  + {{\left| {11} \right\rangle }_{AB}}\left| {{\chi _{0111}}} \right\rangle 
		+ {{\left| {10} \right\rangle }_{BA}}\left| {{\chi _{1000}}} \right\rangle  + {{\left| {11} \right\rangle }_{BA}}\left| {{\chi _{1001}}} \right\rangle \\[2mm]
		+ {{\left| {10} \right\rangle }_{BA}}\left| {{\chi _{1010}}} \right\rangle  + {{\left| {11} \right\rangle }_{BA}}\left| {{\chi _{1011}}} \right\rangle 
		- {{\left| {10} \right\rangle }_{BA}}\left| {{\chi _{1100}}} \right\rangle  - {{\left| {11} \right\rangle }_{BA}}\left| {{\chi _{1101}}} \right\rangle 
		\left. { + {{\left| {10} \right\rangle }_{BA}}\left| {{\chi _{1110}}} \right\rangle  + {{\left| {11} \right\rangle }_{BA}}\left| {{\chi _{1111}}} \right\rangle } \right).
		\label{<12>}%
	\end{array}
\end{align}

If Eve introduces no errors on the SIFT-Z bit, there should be:

\begin{align}
	\begin{array}{c}
		\left| {{\chi _{0000}}} \right\rangle  = \left| {{\chi _{0001}}} \right\rangle  = \left| {{\chi _{0010}}} \right\rangle  = \left| {{\chi _{0011}}} \right\rangle 
		= \left| {{\chi _{0100}}} \right\rangle  = \left| {{\chi _{0101}}} \right\rangle  = \left| {{\chi _{0110}}} \right\rangle  = \left| {{\chi _{0111}}} \right\rangle \\[2mm]
		= \left| {{\chi _{1000}}} \right\rangle  = \left| {{\chi _{1001}}} \right\rangle  = \left| {{\chi _{1010}}} \right\rangle  = \left| {{\chi _{1011}}} \right\rangle 
		= \left| {{\chi _{1100}}} \right\rangle  = \left| {{\chi _{1101}}} \right\rangle  = \left| {{\chi _{1110}}} \right\rangle  = \left| {{\chi _{1111}}} \right\rangle 
		= 0.
	\end{array}
	\label{<13>}%
\end{align}

Therefore, without introducing errors, Eve cannot obtain any information about the key. In the above proof, the analysis of the CTRL bit was not conducted. It can be easily proven that if Eq.\eqref{<13>} holds, then Eq.\eqref{<10>} is also satisfied simultaneously. However, this does not imply that the CTRL operation in this system can be ignored. The proof of the CTRL operation see Appendix~\hyperref[sec:appendixb]{B}. All in all, if Eve attacks the quantum states during the key distribution process to obtain key information, errors would inevitably be introduced into the quantum states of both Alice and Bob, leading to detection. Thus, the design proposed in this paper, based on the “selective modulation”, is completely robust against attacks targeting quantum states. 
\section{Conclusion}\label{sec:6}
We propose a phase-encoding semi-quantum key distribution system based on the Single-state protocol. By applying the “selective modulation" method, classical users do not need to measure or retransmit the photons. We have established a semi-quantum key distribution system with a frequency of 100 MHz. Compared with previous experiments, notable enhancements are observed in experimental metrics, including an interference contrast of 97.45\%, an average quantum bit error rate of 1.20\%, and a raw key rate of 88Kbps, validating the feasibility and stability of our system through experimentation. The classical user of our system requires only two optical devices, significantly reducing equipment requirements and enhancing its potential for widespread application. Furthermore, we have theoretically demonstrated the security of selective modulation against marking attacks, and have shown the robustness of our system in resisting quantum state attacks.
This research is a novel application of the selective modulation method in SQKD and gives a comprehensive theoretical description of the method. Additionally, the classical end design proposed in this paper is simplified, which is expected to fully leverage the potential of SQKD in quantum networks, achieving low-cost and secure key distribution.

%
%

\ack{This work was supported by Natural Science Foundation of Guangdong Province 2024A1515012427, by National Nature Science Foundation of China 62371199, 62071186, and 61771205, and by Guangdong Provincial Key Laboratory of Construction Foundation 2020B1212060066.}



\data{The data that support the findings of this study are available from the corresponding author upon reasonable request.}

\suppdata{
	\section*{Appendix A: Experimental setup details}\label{sec:appendixa}

On Bob's module, the setup comprises four modules: a laser source, an AMZI, phase modulation, and detectors. In the laser source section, laser pulses are generated by a picosecond laser (LD, produced by Qasky, WT-LD100-D). The pulse width of the laser pulses is 50ps, and the system frequency is 100MHz. The isolator (ISO) is used to prevent reflected light from entering the laser. An attenuator (Att) is utilized to attenuate the average photon number per pulse of light pulses to 0.1. The PC and the PMPBS1 are employed to adjust the polarization state of the light pulses and align them with the slow axis of the polarization-maintaining fiber. In the asymmetric Mach-Zehnder interferometer section, the PMCIR is used to transmit the forward light pulses to the 50:50 PMBS and separate the backward light pulses to the detectors. The PMBS, along with PMPBS2, forms the asymmetric Mach-Zehnder interferometer section. The phase modulation section consists of two polarization-independent phase modulators (PM1, PM2, produced by iXblue, MPZ-LN-10-P-P-FA-FA), which modulate the forward and backward light pulses. A signal source (SS1, produced by SIGLENT, SDG6052X-E) drives the phase modulators. The detector section comprises two single-photon detectors (SPD, produced by Qasky, WT-SPD300-ULN), with a detection frequency of 100MHz. SPD1 is connected to the PMCIR port3, while SPD2 is connected to PMBS. The clock synchronization between the laser, detectors, and SS is achieved using a digital generator (DG, produced by Qasky, WT-HGCG200).

On Alice's module, the setup consists of a polarization-independent phase modulator (PM3, produced by KANGGUAN, KG-SM-15-10G-PP-FP) and a Faraday mirror. The phase modulator is driven by SS (SS2, produced by SIGLENT, SDG6052X-E).

In the AMZI section, we introduce a delay of 2.9ns between the two pulses due to the arm length difference. Due to the action of PMPBS, the two light pulses will propagate along the fast axis and slow axis of the polarization-maintaining fiber, respectively. For simplicity, we will refer to them as ${P_F}$ and ${P_L}$. When the pulses arrive at Alice's module, Alice selectively modulates ${P_L}$ using her phase modulator. When the pulses are reflected back to Bob, Bob modulates the pulses using his phase modulator. Due to the Faraday rotation conjugate effect, the polarization of pulse will rotate by $90^{\circ}$. Consequently, both ${P_F}$ and ${P_L}$ will return to the polarization-maintaining beam splitter (PMBS) simultaneously and undergo interference. Ultimately, depending on the phase difference, the pulses will be detected by either SPD1 or SPD2.

The “selective modulation” operation in this paper is described in detail below. Bob sends light pulses. A pulse signal with a repetition frequency of 100MHz and a voltage of 2.95V, generated by SS1, drives the phase modulator PM1. Then the PM1 consistently applies $\frac{\pi }{2}$ phase modulation on ${P_L}$. If Alice chooses CTRL, she does not apply voltage to the phase modulator, and the light pulse will be reflected to Bob directly. If Alice chooses SIFT(0), she applies $\frac{\pi }{2}$ phase modulation on ${P_L}$. In this circumstance, PM3 is driven by a pulse signal with a repetition frequency of 100MHz and a voltage of 2.01V, generated from SS2. If Alice chooses SIFT(1), she applies $ - \frac{\pi }{2}$ phase modulation on ${P_L}$. In this circumstance, PM3 is driven by a pulse signal with a repetition frequency of 100MHz and a voltage of -1.78V (achieved by changing the signal polarity). To maintain the Faraday effect, Alice needs to modulate the forward and backward pulses twice in the same manner, which can be ensured by controlling the length of the fiber between PM3 and FM. Due to the identical modulation twice, the required voltage for applying $\frac{\pi }{2}$ phase modulation is correspondingly reduced.

When the pulses return to Bob, he randomly chooses to measure them in either the Z basis or the X basis. If Bob chooses the Z basis measurement, he drives PM2 with a voltage of 0. If Bob chooses the X measurement, he applies $\frac{\pi }{2}$ phase modulation to ${P_F}$. In this case, PM2 is driven by a pulse signal with a repetition frequency of 100MHz and a voltage of 2.81V, generated by SS2. 

The experiment employed two InGaAs/InP single-photon detectors (produced by Qasky, WT-SPD300-ULN), with a detection gate width of 1ns, a detection frequency of 100MHz, a dark count probability per gate of $3 \times {10^{ - 6}}$, a dead time of 5ns, and an approximate detection efficiency of 20.5\%.

\section*{Appendix B: Proof of the CTRL operation}\label{sec:appendixb}
\setcounter{subsection}{0}
\setcounter{equation}{0}
\renewcommand{\theequation}{B.\arabic{equation}}
Here we will demonstrate the necessity of the CTRL operation for ensuring the security of this system. We will use an example where Eve only attacks the backward channel, which can be seen as a special case of Eve implementing a two-way attack. Since Eve only attacks the backward channel, ${\hat U_F}$ can be regarded as the identity operator ${\hat I_E}$ in the previous analysis. Therefore, the states held by Bob can be represented as:
\begin{align}
	{\left| \psi  \right\rangle _B} = \frac{1}{{\sqrt 2 }}\left( {{{\left| 0 \right\rangle }_B} + {{\left| 1 \right\rangle }_B}} \right).
	\label{<14>}%
\end{align}
When Alice chooses the SIFT(0) operation, the states held by Alice, Bob can be represented as: 
\begin{align}
	{\left| \psi  \right\rangle _{{A_0}B}} = {\left| 0 \right\rangle _A} \otimes {\hat S_0}{\left|  +  \right\rangle _B} = {\left| {00} \right\rangle _{AB}}.
	\label{<15>}%
\end{align}
When Alice chooses the SIFT(1) operation, the states held by Alice, Bob can be represented as: 
\begin{align}
	{\left| \psi  \right\rangle _{{A_1}B}} = {\left| 1 \right\rangle _A} \otimes {\hat S_1}{\left|  +  \right\rangle _B} = {\left| {11} \right\rangle _{AB}}.
	\label{<16>}%
\end{align}
After Alice performs selective modulation, Eve implements an attack ${\hat U_R}$ on the quantum state of the backward channel. The final quantum state of Alice 's chosen CTRL is: 
\begin{align}
	\begin{array}{c}
		{{\hat U}_R}\left( {{{\left| \psi  \right\rangle }_B}{{\left| \chi  \right\rangle }_E}} \right) = \frac{1}{{\sqrt 2 }}\left( {{{\left| 0 \right\rangle }_B}\left| {{\chi _{00}}} \right\rangle } \right.
		\left. { + {{\left| 1 \right\rangle }_B}\left| {{\chi _{01}}} \right\rangle  + {{\left| 0 \right\rangle }_B}\left| {{\chi _{10}}} \right\rangle  + {{\left| 1 \right\rangle }_B}\left| {{\chi _{11}}} \right\rangle } \right).
	\end{array}
	\label{<17>}%
\end{align}
The final quantum state of Alice 's chosen SIFT(0) is:
\begin{align}
	\begin{array}{c}
		{{\hat U}_R}\left( {{{\left| \psi  \right\rangle }_{{A_0}B}}{{\left| \chi  \right\rangle }_E}} \right) = {{\hat U}_R}\left( {{{\left| {00} \right\rangle }_{AB}}{{\left| \chi  \right\rangle }_E}} \right)
		= {\left| {00} \right\rangle _{AB}}\left| {{\chi _{00}}} \right\rangle  + {\left| {01} \right\rangle _{AB}}\left| {{\chi _{01}}} \right\rangle.
	\end{array}
	\label{<18>}%
\end{align}
The final quantum state of Alice 's chosen SIFT(1) is: 
\begin{align}
	\begin{array}{c}
		{{\hat U}_R}\left( {{{\left| \psi  \right\rangle }_{{A_1}B}}{{\left| \chi  \right\rangle }_E}} \right) = {{\hat U}_R}{\left| {11} \right\rangle _{AB}}{\left| \chi  \right\rangle _E}
		= {\left| {10} \right\rangle _{AB}}\left| {{\chi _{10}}} \right\rangle  + {\left| {11} \right\rangle _{AB}}\left| {{\chi _{11}}} \right\rangle.
	\end{array}
	\label{<19>}%
\end{align}
If Eve introduces no errors on the SIFT-Z bit, there should be:
\begin{align}
	\left| {{\chi _{01}}} \right\rangle  = \left| {{\chi _{10}}} \right\rangle  = 0.
	\label{<20>}%
\end{align}
Then the Eq. \ref{<17>} can be rewritten as: 
\begin{align}
	{\hat U_R}\left( {{{\left| \psi  \right\rangle }_B}{{\left| \chi  \right\rangle }_E}} \right) = \frac{1}{{\sqrt 2 }}\left( {{{\left| 0 \right\rangle }_B}\left| {{\chi _{00}}} \right\rangle  + {{\left| 1 \right\rangle }_B}\left| {{\chi _{11}}} \right\rangle } \right) .
	\label{<21>}%
\end{align}
If Eve introduces no errors on the CTRL-X bit, there should be: 
\begin{align}
	\left| {{\chi _{00}}} \right\rangle  = \left| {{\chi _{11}}} \right\rangle .
	\label{<22>}%
\end{align}

Therefore, without introducing errors, Eve cannot obtain any information about the key. The preceding discussion highlights the necessity of the classical CTRL operation to ensure the security of SQKD. 

}


\bibliographystyle{iopart-num.bst}
\bibliography{article}

\end{document}